\documentclass[a4paper,11pt]{article}
\usepackage{jheppub,braket} 

\title{Fluctuation in the Fidelity of Information Recovery from Hawking Radiation}

\author{Masamichi Miyaji$^{1,2}$, Kazuyoshi Yano$^{2}$}

\affiliation{$^1$ Institute for Advanced Research, Nagoya University, Nagoya, Aichi 464-8601, Japan}
\affiliation{$^2$Department of Physics, Nagoya University, Nagoya, Aichi 464-8602, Japan}
\emailAdd{miyaji.masamichi.j4@f.mail.nagoya-u.ac.jp, kazuyoshi@eken.phys.nagoya-u.ac.jp}

\abstract{ The interior of a pure-state black hole is known to be reconstructed from the Petz map by collecting a sufficiently large amount of the emitted Hawking radiation. This was established based on the Euclidean replica wormhole, which comes from an ensemble averaging over gravitational theories. On the other hand, this means that the Page curve and the interior reconstruction are both ensemble averages; thus, there is a possibility of large errors. In the previous study \cite{Bousso:2023efc}, it was shown that the entropy of the Hawking radiation has fluctuation of order $e^{-S_{\mathbf{BH}}}$, thus is typical in the ensemble. In the present article, we show that the fluctuations of the relative entropy difference in the encoding map and the entanglement fidelity of the Petz map are both suppressed by $e^{-S_{\mathbf{BH}}}$ compared to the signals, establishing the typicality in the ensemble. In addition, we also compute the entanglement loss of the encoding map.}

\numberwithin{equation}{section}

\begin{document}
\maketitle
\flushbottom

\section{Introduction}

In AdS/CFT, the entanglement entropy of a boundary subregion $A$ is given by the area of the Ryu-Takayanagi surface $\gamma_A$, called Ryu-Takayanagi formula \cite{Ryu:2006bv, Hubeny:2007xt, Faulkner:2013ana, Engelhardt:2014gca}. The bulk subregion $EW[A]$ bounded by $A$ and $\gamma_A$, is called the entanglement wedge of $A$, and it corresponds to the boundary subregion $A$ in the following sense: the information on $EW[A]$ can be embedded in the boundary subregion $A$, and there is an approximate inverse 
which reconstructs the original bulk information. This whole process is called entanglement wedge reconstruction \cite{Czech:2012bh, Almheiri:2014lwa, Pastawski:2015qua, Jafferis:2015del, Dong:2016eik,Hayden:2016cfa, Harlow:2016vwg, Cotler:2017erl, Chen:2019gbt}. 

This existence of an approximate recovery map is crucial in the recovery of the bulk information from the boundary state. It is guaranteed due to the fact that the \emph{relative entropy difference} between the bulk and the boundary states is sufficiently small \cite{Jafferis:2015del}, which means the bulk information is faithfully encoded in the boundary. When there is an approximate recovery map, then the \emph{Petz map} is known to be an approximate inverse map whose error is in the same order \cite{Barnum}, and thus, we can use the Petz map to reconstruct interior information from the boundary. This means, in general situations, it is sufficient to study the \emph{fidelity of the Petz map} when we need to check if there is an approximate recovery map or not. 

It was noticed later that the entanglement wedge reconstruction can be applied to retrieving the interior information of the black hole \cite{Penington:2019npb, Almheiri:2019psf,Almheiri:IslandFormula, Penington:ReplicaWormholeWestCoast, Almheiri:ReplicaWormholeEastCoast}, see also \cite{Vardhan:2021mdy,Balasubramanian:2022fiy,Akers:2022qdl,Czech:2023rbh, Nakayama:2023kgr} for recent treatments. It was found that the information inside the black hole horizon sufficiently after the Page time is encoded in the Hawking radiation and, thus, can be recovered using the entanglement wedge reconstruction from the Hawking radiation. The earlier special version of this reconstruction is the Hayden-Preskill protocol \cite{Hayden:2007cs}, where the size of the interior information remains $O(1)$ \cite{Hayden:2018khn}. In the present case, it was found in \cite{Penington:ReplicaWormholeWestCoast} that the interior information, whose dimension scales as the dimension of the black hole, can be recovered via the Petz map in accord with \cite{Hayden:2018khn}. 

In the derivations of the Page curve and of the reconstruction of the interior information, the Euclidean replica wormholes played a crucial role. These Euclidean wormholes were found to bring a new important puzzle since the gravitational path-integral over Euclidean wormholes cannot be dual to quantum mechanics, but instead dual to an ensemble of boundary quantum theories \cite{Maldacena:2004rf, Maldacena:2004rf, Saad:2018bqo, Saad:2019pqd, Harlow:2018tqv, Bousso:2019ykv, Bousso:2020kmy}. This implies that the gravity theory with Euclidean wormholes is also an ensemble of quantum gravity theories. In the ensemble of theories, the physical quantities are all ensemble averages. Thus, the computed Page curve and the fidelity of the Petz map are both averages in the ensemble, and there is no guarantee that they are typical answers in the ensemble. In other words, it is possible that these quantities are significantly different from the averaged answers for typical theories in the ensemble. A famous example of such a large deviation can be seen in the ramp and the plateau of the spectral form factor, where the error and the average answer are in the same order. Therefore, the fundamental question to ask in the black hole information problem is whether the ensemble averages of these quantities are typical or not in the ensemble.

In \cite{Bousso:2023efc}, this question on the typicality of the Page curve was addressed. It was shown that the average noise in the entropy of the Hawking radiation is always of order $e^{-S_{\text{BH}}}$, where $S_{\text{BH}}$ is the black hole entropy. This means that the fluctuation is proportional to the inverse of the Hilbert space dimension. Therefore, the ensemble average of the Page curve is highly typical in the ensemble.

In this paper, we ask another important question in the typicality, namely that of interior information recovery. We will show that the reconstruction of the interior information after the Page time is possible in the typical member of the ensemble. We show this by studying the fluctuations of the measures of the information recovery, namely the relative entropy difference and the entanglement fidelity of the Petz recovery map. When the relative entropy difference is sufficiently small or the entanglement fidelity is close to one, we can tell that the information recovery from the interior is possible. These measures are known to take those values in average sufficiently after the Page time, but so far the typicality of that behavior has not been discussed. We found that such behavior is indeed typical in the ensemble, namely, if we take a member in the ensemble, then the information recovery is typically possible. We show this by studying the fluctuations of these measures and find that they are suppressed by the factor $e^{-S_{\text{BH}}}$ compared to the signals. This implies that the success of the information recovery is also highly typical in the ensemble.

The organization of this paper is as follows. In section \ref{section:QI}, we explain fundamental notions and measures of the interior information recovery, such as the relative entropy difference, the entanglement fidelity, the Petz map, and the coherent quantum information. We will explain why these are so important in the reconstruction by explaining the fundamental inequalities they satisfy. In section \ref{section:PSSY}, we review the PSSY model, which is the model of a black hole entangled with its Hawking radiation. We will also review the fluctuation of the Page curve in the PSSY model studied in \cite{Bousso:2023efc}. In section \ref{section:recovery}, we present our main result of this paper, namely, we evaluate the fluctuations of the relative entropy difference and the entanglement fidelity of the Petz map.


\section{Measures of the Interior Information Recovery}\label{section:QI}
In this section, we will review some important notions and measures in interior information recovery. We will briefly explain the quantum error correction and the Petz map, which are nothing but the entanglement wedge reconstruction and the bulk recovery map when applied to gravity. We then explain how it can be studied using the relative entropy difference, the entanglement fidelity, and the coherent quantum information. Readers who are familiar with their properties can skip this section.

\subsection{Quantum Error Correction and the Petz Map}
In quantum error correction, we consider embedding quantum information into a larger system to protect the information against possible errors. In the entanglement wedge reconstruction of the black hole interior, the quantum information of the interior excitations is embedded into the Hawking radiation. The bulk interior quantum information lives in a Hilbert space $H_{\text{code}}$ as a density matrix on $H_{\text{code}}$. This Hilbert space $H_{\text{code}}$ is called {\it{code subspace}}, and the space of quantum states on $H_{\text{code}}$ is denoted as $\mathcal{S}(H_{\text{code}})$. Now, the quantum information on $H_{\text{code}}$ is embedded into the larger one $H_{\bold{R}}$. $H_{\bold{R}}$ corresponds to the Hilbert space of the Hawking radiation. The embedding map $\mathcal{N}:\mathcal{S}(H_{\text{code}})\rightarrow\mathcal{S}(H_{\bold{R}})$ is called the encoding map. The encoding map is a completely positive trace preserving (CPTP) map, meaning that any entangled quantum state should be mapped to an entangled quantum state. We will call a CPTP map as a channel as well. 

In order to recover the interior information from the Hawking radiation, we need to find a CPTP map $\mathcal{R}: \mathcal{S}(H_{\bold{R}})\rightarrow\mathcal{S}(H_{\text{code}})$ that satisfies
\begin{equation}
    (\mathcal{R}\circ\mathcal{N})|_{\mathcal{S}(H_{\text{code}})}\approx \mathbb{I}_{\mathcal{S}(H_{\text{code}})},
\end{equation}
so that the embedded information can be reconstructed using $\mathcal{R}$. Such map $\mathcal{R}$ is called an approximate recovery map for $\mathcal{N}$. As we will explain shortly, it is known that the Petz map can be used as a good approximate recovery map whenever there is an approximate recovery map. Thus, we can use the Petz map as an explicit realization of the approximate recovery map. To write down the Petz map, we need to fix a $\rho_0\in \mathcal{S}(H_{\text{code}})$, and the Petz map depends on $\rho_0$. The Petz map is explicitly given by
\begin{equation}
    \mathcal{R}^{\text{Petz}}_{\mathcal{N},\rho_0}[\sigma]:=\rho_0^{1/2}\mathcal{N}^{\dagger}[\mathcal{N}[\rho_0]^{-1/2}\sigma\mathcal{N}[\rho_0]^{-1/2}]\rho_0^{1/2}.
\end{equation}
 Here $\mathcal{N}^{\dagger}$ is the adjoint defined by $\text{Tr}[\sigma^{\dagger}\mathcal{N}[\rho]]=\text{Tr}[\mathcal{N}^{\dagger}[\sigma]^{\dagger}\rho]$. It is clear that the Petz map recovers the reference state, namely $\mathcal{R}^{\text{Petz}}_{\mathcal{N},\rho_0}[\mathcal{N}[\rho_0]]=\rho_0$ holds.

\subsection{Entanglement Fidelity and the Petz Map}

The fidelity of the channel is widely used in quantum information theory in order to measure the difference between channels. In particular, we will be interested in the difference between $(\mathcal{R}\circ\mathcal{N})|_{\mathcal{S}(H_{\text{code}})}$ and $\mathbb{I}_{\mathcal{S}(H_{\text{code}})}$, in order to measure how well $\mathcal{R}$ can recover the original information encoded by $\mathcal{N}$. For this purpose, we consider the \emph{entanglement fidelity} of a channel $\mathcal{T}:S(H_{\text{code}})\rightarrow S(H_{\text{code}})$ of a state $\rho\in S(H_{\text{code}})$, defined by
\begin{eqnarray}
    F_e(\rho,~\mathcal{T})&:=&\langle\Phi_{\rho}|(\mathcal{T}\otimes\mathbb{I}_{H'})(|\Phi_\rho\rangle\langle\Phi_\rho|)|\Phi_{\rho}\rangle,
\end{eqnarray}
here we used arbitrary purification $|\Phi_{\rho}\rangle\in H_{\text{code}}\otimes H'$ of $\rho$. Note that this definition does not depend on the choice of the purification. The entanglement fidelity depends on the state $\rho$ and does not capture the worst case deviation from the identity channel, in contrast to the minimum fidelity \footnote{The \emph{minimum fidelity} $F(\mathcal{T})$ measures the worst case deviation of the channel $\mathcal{T}$ from the identity channel $\mathbb{I}$, and is defined by
\begin{eqnarray}
    F_{min}(\mathcal{T}):=\underset{|\phi\rangle\in H,~\langle\phi|\phi\rangle=1}{\text{Min}}\langle\phi|\mathcal{T}(|\phi\rangle\langle\phi|)|\phi\rangle.
\end{eqnarray}
By definition, it is clear that 
\begin{equation}
    1\geq F_e(\rho,~\mathcal{T})\geq F_{min}(\mathcal{T}),
\end{equation}
and $F_{min}(\mathcal{N})=1$ iff $\mathcal{T}=\mathbb{I}$.

The entanglement fidelity gives a lower bound on the minimum fidelity when the Hilbert space is appropriately restricted. More explicitly, it can be shown that for each $k$ there exist a $k$-dimensional subspace $H_k\subset H$ and associated projection $P_k$, such that for the restricted channel 
\begin{equation}
    \mathcal{T}_k(\sigma):=P_k\mathcal{N}(\sigma)P_k+\text{Tr}\Big[(1-P_k)\mathcal{T}(\sigma)\Big]k^{-1}P_k,
\end{equation}
the inequality
\begin{equation}
    F(\mathcal{T}_k)_{H_k}\geq 1-\frac{1-F_e(\rho,~\mathcal{T})}{1-k||\rho||_{\infty}},
\end{equation}
holds \cite{Kretschmann_2004, 5429118}. Here we used the standard p-norm $||\rho||_{p}:=(\text{Tr}[\rho^p])^{1/p}$. If we take the maximally mixed state $\rho_0=\mathbb{I}_H/\dim H$ and $k=\dim H/2$ in this theorem, we obtain 
\begin{equation}
    F(\mathcal{T}_k)_{H_k}\geq 1-2(1-F_e(\rho,~\mathcal{T})).
\end{equation}
Thus, once we show the entanglement fidelity is sufficiently close to $1$, then we can automatically see the fidelity on a restricted subspace is also close to $1$. The important point here is that there is no dimension-dependent factor in this inequality, which would prohibit us from considering large code subspace dimension, for instance.}. By definition, it is clear that. 
\begin{equation}
    1\geq F_e(\rho,~\mathcal{T})\geq 0,
\end{equation}
and $F_e(\rho,~\mathcal{T})=1$ iff $\mathcal{T}|_{\text{supp}(\rho)}=\mathbb{I}|_{\text{supp}(\rho)}$. Thus, we can see that the entanglement fidelity measures the closeness of $\mathcal{T}$ to the identity on the support of $\rho$.

Now, we use the entanglement fidelity to measure how well the Petz map $\mathcal{R}^{\text{Petz}}_{\mathcal{N},\rho}$ recovers the original information, compared to arbitrary channel $\mathcal{R}$. Indeed, if the map $\mathcal{R}$ is an approximate recovery map, then the Petz map is known to be an equally good approximate recovery map shown in \cite{Barnum}. Namely, there is an inequality
\begin{equation}
     F_e(\rho,\mathcal{R}^{\text{Petz}}_{\mathcal{N},\rho}\circ\mathcal{N})\geq F_e(\rho,\mathcal{R}\circ\mathcal{N})^2.
\end{equation}
This implies, if $F_e(\rho,\mathcal{R}\circ\mathcal{N})=1-\epsilon$, then $F_e(\rho,\mathcal{R}^{\text{Petz}}_{\mathcal{N},\rho}\circ\mathcal{N})\geq 1-2\epsilon$. Thus, the Petz map is always as good as the best approximate recovery map in terms of entanglement fidelity. In the black hole interior reconstruction, we will use the entanglement fidelity to see if the Petz map can indeed recover the interior information.

\subsection{Reversibility and the Relative Entropy Difference}

Without using the explicit Petz map, we can tell whether there is an approximate recovery map or not by considering reversibility measures. One such measure is the relative entropy difference. Consider quantum states $\rho,~\sigma\in \mathcal{S}(H_{\text{code}})$ and the encoding map $\mathcal{N}:\mathcal{S}(H_{\text{code}})\rightarrow\mathcal{S}(H_{\text{R}})$. The relative entropy $D(\sigma|\rho)$ is defined by
\begin{equation}
    D(\sigma|\rho)=\text{Tr}[\sigma(\log\sigma-\log\rho)],
\end{equation}
which measures the difference between the two states. The relative entropy difference is 
\begin{equation}
    D(\sigma|\rho)-D(\mathcal{N}[\sigma]|\mathcal{N}[\rho]),
\end{equation}
which is guaranteed to be non-negative due to the monotonicity of the relative entropy. When the encoding map $\mathcal{N}$ is erroneous, then the difference between states $\mathcal{N}[\sigma]$ and $\mathcal{N}[\rho]$ will be lost, resulting in an increase of the relative entropy difference. 

Conversely, if the relative entropy difference is small enough, then it is known that the averaged rotated Petz map, which only depends on the reference state $\rho_0$, is an approximate recovery map for $\mathcal{N}$. Indeed, it is known that the averaged rotated Petz map satisfies, for all $\sigma\in\mathcal{S}(H_{\text{code}})$ satisfying $\text{supp}(\sigma)\subset\text{supp}(\rho_0)$  \cite{Junge:2015lmb}
\begin{equation}\label{eq:rotated}
    D(\sigma|\rho_0)-D(\mathcal{N}[\sigma]|\mathcal{N}[\rho_0])\geq
    -2\log F\left((\mathcal{R}^{\text{Averaged Rotated Petz}}_{\mathcal{N},\rho_0}\circ\mathcal{N})[\sigma]\Big{|}\sigma\right).
\end{equation}
Here the fidelity $F(\rho|\sigma)$ is defined by
\begin{equation}
    F(\rho|\sigma):=||\sqrt{\rho}\sqrt{\sigma}||_1=\underset{|\Phi_\rho\rangle,|\Phi_{\rho}\rangle}{\text{Max}}|\langle\Phi_{\rho}|\Phi_{\sigma}\rangle|,
\end{equation}
where $|\Phi_\rho\rangle,|\Phi_{\sigma}\rangle\in H_{\text{code}}\otimes H'$ are arbitrary purifications of $\rho, \sigma\in S(H_{\text{code}})$. It is clear that $1\geq F(\rho|\sigma)\geq 0$, and $F(\rho|\sigma)=1$ iff $\rho=\sigma$. The fidelity measures the difference between quantum states. Thus (\ref{eq:rotated}) implies that when the relative entropy is small, the fidelity is close to one, ensuring the information recovery.
\begin{equation}
    (\mathcal{R}^{\text{Averaged Rotated Petz}}_{\mathcal{N},\rho_0}\circ\mathcal{N})[\sigma]\approx\sigma.
\end{equation}
In the black hole interior reconstruction, we will use the relative entropy difference to see if a map that reconstructs the interior information exists.

\subsection{Reversibility and the Entanglement Loss}

Next, we describe another reversibility measure called \emph{coherent information loss}. When the coherent information loss is small, it is guaranteed that there is an approximate recovery map. The result in this subsection will be used in subsection \ref{section:coherent}.

We first describe the coherent information loss. For quantum state $\rho\in\mathcal{S}(H_{\text{code}})$ and the encoding map $\mathcal{N}:\mathcal{S}(H_{\text{code}})\rightarrow\mathcal{S}(H_{\bold{R}})$, the coherent information is defined by
\begin{equation}
    I_c(\rho,\mathcal{N}):=S(\mathcal{N}[\rho])-S((\mathcal{N}\otimes\mathbb{I}_{H'})[|\Phi_{\rho}\rangle\langle\Phi_{\rho}|])
\end{equation}
here we consider an arbitrary purification $|\Phi_{\rho}\rangle\in H_{\text{code}}\otimes H'$ of $\rho$. 
Then coherent information loss is defined by
\begin{eqnarray}
    \delta_c(\rho,\mathcal{N})&:=&S(\rho)-I_c(\rho,\mathcal{N}).
\end{eqnarray}
The coherent information loss measures the information loss in the encoding map. Indeed, if the coherent information loss is small, then there is a map $\mathcal{R}:\mathcal{S}(H_{\bold{R}})\rightarrow\mathcal{S}(H_{\text{code}})$ which recovers the original information. Indeed, the entanglement fidelity is lower bounded as \cite{schumacher2002} 
\begin{equation}\label{eq:ineq}
    \underset{\mathcal{R}}{\text{Max}}~F_e(\rho,\mathcal{R}\circ\mathcal{N})\geq 1-\sqrt{2\delta_c(\rho,\mathcal{N})}.
\end{equation}
Thus, if the coherent information loss is sufficiently small, then there exists a map $\mathcal{R}$ such that $F_e(\rho,\mathcal{R}\circ\mathcal{N})\approx 1$. In other words, $\mathcal{R}$ is an approximate recovery map in terms of the entanglement fidelity when the coherent information loss is sufficiently small \footnote{This existence criterion for the recovery channel is equivalent to the \emph{decoupling criterion}. To see this, let us use the Stinespring dilatation and express the channel $\mathcal{N}$ as a unitary map $U$ on an extended Hilbert space $H_{\text{code}}\otimes H_E=H_{\bold{R}}\otimes H_{E'}$ as
\begin{equation}
    \mathcal{N}[\rho]
    =\text{Tr}_{H_{E'}}\left[U(\rho\otimes|0\rangle\langle0|_{H_E})U^{\dagger}\right].
\end{equation}
We then consider the purification $|\Phi_{\rho}\rangle\in H_{\text{code}}\otimes H'$. Then we obtain the pure state $|\Psi_{\rho,U}\rangle:=(U\otimes\mathbb{I}_{H'})|\Phi_{\rho}\rangle\otimes|0\rangle_{H_{E}}\in H_{\bold{R}}\otimes H_{E'}\otimes H_{H'}$. Using the quantum state $\rho_{H'H_{E'}}:=\text{Tr}_{H_{\bold{R}}}[|\Psi_{\rho,U}\rangle\langle\Psi_{\rho,U}|]$ on $H'\otimes H_{E'}$, the coherent information loss is rewritten as
\begin{equation}
    \delta_c(\rho,\mathcal{N})=D(\rho^{\mathcal{N}}_{H'H_{E'}}|\rho^{\mathcal{N}}_{H'}\otimes\rho^{\mathcal{N}}_{H_{E'}}).
\end{equation}
Thus vanishing coherent information loss is equivalent to the factorization
\begin{equation}\label{eq:fact}
    \rho_{H'H_{E'}}=\rho_{H'}\otimes\rho_{H_{E'}}.
\end{equation}
Thus, the factorization (\ref{eq:fact}) is equivalent to the existence of the exact recovery map, and this condition (\ref{eq:fact}) for the information recovery is called the decoupling criterion.}.

The coherent information can be replaced by other correlation measures, such as mutual information, or entanglement measures like squashed entanglement. For the mutual information loss $\delta_{m}(\rho,\mathcal{N}):=S(\rho)-I(H':H_{\bold{R}})/2$, similar to the coherent information loss, we have \cite{PhysRevA.77.012309, FB}
\begin{eqnarray}\label{eq:mutual}
    \underset{\mathcal{R}}{\text{Max}}~F_e(\rho,\mathcal{R}\circ\mathcal{N})
    &\geq& 
    1-2\sqrt{\delta_{m}(\rho,\mathcal{N})}.
\end{eqnarray}
Thus, when the mutual information loss is small, there exists an approximate recovery map $\mathcal{R}$.


\section{The PSSY Model}\label{section:PSSY}

In the present section, we review the PSSY model \cite{Penington:ReplicaWormholeWestCoast}, which is a toy model of the black hole entangled with Hawking radiation. The model consists of the JT gravity \cite{Teitelboim:1983ux, Jackiw:1984je, Maldacena:2016upp, Stanford:2017thb, Yang:2018gdb, Saad:2019lba,Stanford:2019vob, Saad:2019pqd} with an end-of-the-world (EOW) brane in bulk anchored at the boundary. The EOW brane has $k$ flavors and tension $\mu~(\geq0)$. The action is
\begin{equation}
    S=S_{\text{JT}}+S_{\text{Brane}},
\end{equation}
where
\begin{equation}
    S_{\text{JT}}=-\frac{S_0}{4\pi}\left(\int_{\mathcal{M}} \sqrt{g}R+2\int_{\partial\mathcal{M}}\sqrt{h}K\right)
    -
    \frac{1}{2}\left(\int_{\mathcal{M}} \sqrt{g}\phi(R+2)+\int_{\partial\mathcal{M}}\sqrt{h}\phi K\right),
\end{equation}
\begin{equation}
    S_{\text{brane}}=\mu\int_{\text{Brane}}ds,
\end{equation}
with the standard asymptotic boundary condition
\begin{equation}
    ds^2|_{\partial{\mathcal{M}}}=\frac{d\tau^2}{\epsilon^2},~\phi|_{\partial{\mathcal{M}}}=\frac{1}{\epsilon},\end{equation}
with $\tau$ being the boundary Euclidean time. When analytically continued to Lorentzian time, the EOW brane plays the role of a particle with a flavor behind the black hole horizon. In terms of boundary quantum mechanics, $|i\rangle_{\mathbf{EOW}}$ is a random sum over energy eigenstates $|E_s\rangle$ of a  Hamiltonian in the matrix integral dual to the JT gravity \cite{Penington:ReplicaWormholeWestCoast}
\begin{equation}\label{eq:bulkstate}
    |i\rangle_{\bold{EOW}}
    \propto\sum_{s}\sqrt{f(E_s)}2^{1/2-\mu}\Gamma[\mu-1/2+i\sqrt{2E_s}]C_{(ia)s}|E_s\rangle.
\end{equation}
Here $C_{is}$ is a complex Gaussian random matrix of $k\times \infty$ entries, and $f(E)$ is a continuous function of $E$, which defines the ensemble of states such as a microcanonical ensemble. EOW brane states $|i\rangle_{\mathbf{EOW}}$ with different flavors are nearly orthogonal but have a small overlap of order $e^{-S_0/2}$. This small overlap accounts for the finiteness of the gravity Hilbert space dimension $e^{S_0}$. The accumulation of these small overlaps results in the unitary Page curve of the Hawking radiation. 

In the PSSY model, one considers a state in which the EOW brane flavor is maximally entangled with an auxiliary non-gravitating system $\mathbf{R}$
\begin{equation}\label{eq:state}
    |\Psi_a\rangle=\frac{\sum_{i=1}^k|i\rangle_{\mathbf{R}}|i\rangle_{\mathbf{EOW}}}{\sqrt{\sum_i\langle i|i\rangle_{\mathbf{EOW}}}}.
\end{equation}
Note that the state (\ref{eq:state}) is always normalized. The reduced density matrix of the radiation is
\begin{equation}\label{eq:reduceddensity}
    \hat{\rho}_{\mathbf{R}}=\frac{\sum_{i,j=1}^k|i\rangle\langle j|_{\mathbf{R}}\langle j|i\rangle_{\mathbf{EOW}}}{\sum_i\langle i|i\rangle_{\mathbf{EOW}}}~.
\end{equation}

\subsection{JT Gravity Partition Functions}

For reference, we briefly review the JT gravity partition functions in a microcanonical ensemble with EOW branes. For full details, we refer to \cite{Bousso:2023efc}. We first introduce a continuous smearing function $f(x)$ to define a microcanonical ensemble. The smearing function $f(x)$ is defined as 
\begin{equation}
    f(x)=\left\{
    \begin{aligned}
    &1&  (|x-E|&< \Delta E/2-a/2)\\
    &1-\frac{|x-E|-(\Delta E/2-a/2)}{a/2}&  ( \Delta E/2-a/2&<|x-E|<\Delta E/2)\\
    &0&  ( & \text{otherwise})
    \end{aligned}
    \right.
\end{equation}
This smearing is introduced in order to suppress contributions from higher genus contributions in the fluctuations. We note that this smearing is unnecessary for the evaluations of Renyi entropy, relative entropy, and Petz fidelity.

We first consider the disk partition function. The normalized density of states of the disk is
\begin{equation}
    D_{\text{Disk}}(E)=\frac{\sinh(2\pi\sqrt{2E})}{2\pi^2}.
\end{equation}
Replacing the $n$ geodesic boundaries by $n$ EOW branes with action $S_{\text{EOW}_i}=\mu l_i$, we obtain the bulk partition function
\begin{eqnarray}
    Z^{(n)}_{\text{Disk}}\left[\text{canonical}, \text{boundary lengths}=x_i\right]=e^{S_0}\int_0^{\infty}dED_{\text{Disk}}(E)h(E,\mu)^ne^{-(x_1+....+x_n)E},\nonumber\\
\end{eqnarray}
where we define
\begin{eqnarray}
    h(E,\mu)=\frac{|\Gamma(\mu-1/2+i\sqrt{2E})|^2}{2^{2\mu-1}},
\end{eqnarray}
for $\text{Re}[\mu]-\frac{1}{2}-|\text{Im}[\sqrt{8E}]|>0$.
Thus, the microcanonical partition function with $n$ boundaries is given by
\begin{eqnarray}
    Z^{(n)}_{\text{Disk}}
    \left[\text{microcanonical}, \text{energy}=E, \text{width}=\Delta E\right]    \approx e^{S(E)}h(E,\mu)^{n}.
\end{eqnarray}
Here we denote
\begin{equation}
    e^{S(E)}:=e^{S_0}D_{\text{Disk}}(E)\Delta E,
\end{equation}
which is the number of states in the microcanonical window. Next, we consider geometries with a single double trumpet. Defining
\begin{equation}
    D_{\text{Double Trumpet}}(E,E'):=-\frac{E+E'}{4\pi^2\sqrt{EE'}(E-E')^2},
\end{equation}
the partition function of a single tubular wormhole exchange between two disk topology $n$- and $m$-boundary partition functions is given by
\begin{eqnarray}
    &&Z^{(n,m)}_{\text{Double Trumpet}}(x_1,...,x_n:y_1,...,y_m)
    \nonumber\\&=&
    \int_0^{\infty}dEdE'D_{\text{Double Trumpet}}(E,E')e^{-(x_1+...)E-(y_1+...)E'}
    h(E,\mu)^nh(E',\mu)^m.
\end{eqnarray}
By considering microcanonical ensemble with the smearing function, we arrive at
\begin{eqnarray}\label{eq:doubletrumpet}
    \label{eq:ramp1}Z^{(n,m)}_{\text{Double Trumpet}}
    \left[\text{microcanonical}, \text{energy}=E, \text{width}=\Delta E\right]    \approx\frac{\log (e^{\frac{3}{2}}\frac{\Delta E}{a})}{\pi^2}h(E,\mu)^{n+m}.\nonumber\\
\end{eqnarray}
In the following, we will always consider microcanonical ensemble with smearing function $f(x)$ and take large $k$ and $e^{S_0}$ limit. 

\subsection{Entropy and its Fluctuation}

Utilizing the partition function of the JT gravity with EOW branes in the last subsection, the Renyi entropy of the Hawking radiation was evaluated. Then, it is possible to evaluate the entropy of the Hawking radiation \cite{Penington:ReplicaWormholeWestCoast} and its fluctuation \cite{Bousso:2023efc}. From the results, we can conclude that the entropy and the rank of the Hawking radiation are typical in the ensemble. 

The entropy of the Hawking radiation is
\begin{equation}\label{eq:Pagecurve}
    S_{\mathbf{R}}=\left\{
    \begin{aligned}
    & \log k-\frac{k}{2e^{S(E)}}+O(k^{-1})&  (k&< e^{S(E)})\\
    & S(E)-\frac{e^{S(E)}}{2k}+O(e^{-S(E)})&  (k&> e^{S(E)}),
    \end{aligned}
    \right.
\end{equation}
and the rank of the Hawking radiation state is given by
\begin{equation}
    \text{Tr}[\rho_{\mathbf{R}}^{0+}]=\left\{
    \begin{aligned}
    &k+O(1) &  (k&<e^{S(E)})\\
    &e^{S(E)}+O(1) &  (k&>e^{S(E)}).
    \end{aligned}
    \right.
\end{equation}
The fluctuation of the entropy is given by
\begin{eqnarray}\label{eq:entropy}
    \delta S_{\mathbf{R}}=e^{-S(E)}\times\left\{
    \begin{aligned}
    &\sqrt{\frac{1}{2}-\frac{k}{4e^{S(E)}}+\frac{\log (e^{\frac{3}{2}}\frac{\Delta E}{a})}{4\pi^2}\frac{k^2}{e^{2S(E)}}}+O(e^{2S(E)}k^{-3})&  &(1\ll e^{S(E)}-k)\\
    &\sqrt{\frac{e^{2S(E)}}{2k^2}-\frac{e^{3S(E)}}{4k^3}+\frac{\log (e^{\frac{3}{2}}\frac{\Delta E}{a})}{\pi^2}\left(1-\frac{e^{S(E)}}{2k}\right)^2}+O(e^{-S(E)})&  &(k-e^{S(E)}\gg1).
    \end{aligned}
    \right.
\end{eqnarray}
Thus, the entropy fluctuation is always of order $e^{-S(E)}$. This behavior is distinct from the random pure state, whose entropy fluctuation decays as $k^{-1}$ at large $k$. We can understand this behavior by noticing the fluctuation of the number of states in the given microcanonical window, which is given by the rank fluctuation at large $k$. The fluctuation of the rank is
\begin{equation}\label{eq:summary2}
    \delta\text{Tr}[\hat{\rho}_{\mathbf{R}}^{0+}]
    =\left\{
    \begin{aligned}
    &~\,0 ~~ +O\left(k^{-1}\right)&  &(1\ll e^{S(E)}-k)\\
    & \frac{1}{\pi}\sqrt{\log(e^{\frac{3}{2}}\frac{\Delta E}{a})}+O\left(e^{-S(E)}\right)&  &(k-e^{S(E)}\gg1~),
    \end{aligned}
    \right.
\end{equation}
which is non-zero only after the Page time $k=e^{S(E)}$. Thus, the square of the fluctuation of the number of states depends logarithmically on the ratio $\Delta E/a$.


\section{Fluctuations in the Measures of the Interior Information Recovery} \label{section:recovery}

In this section, we consider the reconstruction of the interior information using the Petz map. It was found that by applying the Petz map to the Hawking radiation after the Page time, the interior information can be reconstructed\cite{Penington:ReplicaWormholeWestCoast}, see also \cite{Hayden:2007cs, Hayden:2018khn}. The recovery of the interior information can be confirmed by the relative entropy difference, the entanglement fidelity of the Petz map, and the coherent information loss. 

The primary goal of this paper is to study the fluctuations of these measures of interior information recovery, in particular, the relative entropy difference and the entanglement fidelity of the Petz map. By studying these fluctuations, we reveal that interior reconstruction is indeed possible among the typical members of the ensemble. We also present a computation of the coherent information loss and the mutual information loss, which guarantee that the information recovery is possible on average.

This section consists of four subsections. We first review how the interior information is encoded in the Hawking radiation state. Then, we study the fluctuations of the relative entropy difference and the entanglement fidelity of the Petz map. Finally, we attach a computation of the measures of the entanglement loss.

\subsection{Encoding Information in the PSSY Model}
We now review how to encode interior quantum information into the Hawking radiation in the PSSY model. For this purpose we generalize the total pure state (\ref{eq:state}) to have indices from the code subspace $|a\rangle\in H_{\text{code}}:=\mathbb{C}^d$
\begin{equation}
    |\Psi_a\rangle=k^{-1/2}\sum_{i=1}^k|i\rangle_{\mathbf{R}}|(ia)\rangle_{\mathbf{EOW}}\in H_{\bold{R}}\otimes H_{\bold{EOW}}.
\end{equation}
Here the EOW brane states have indices $(ia)$. The boundary description of this state $|\Psi_a\rangle$ can be obtained by replacing $i$ by $(ia)$ in (\ref{eq:bulkstate}). These new indices now carry interior information. A quantum state $\rho=\sum_{ab}c_{ab}|a\rangle\langle b|\in S(H_{\text{code}})$ on the code subspace is now encoded in the Hawking radiation via the encoding map $\mathcal{N}$ defined by
\begin{equation}
    \mathcal{N}\left[\sum_{ab}c_{ab}|a\rangle\langle b|\right]
    =
    \text{Tr}_{\mathbf{EOW}}\left[\sum_{ab}c_{ab}\frac{|\Psi_a\rangle\langle\Psi_b|}{\sqrt{\langle\Psi_a|\Psi_a\rangle\langle\Psi_b|\Psi_b\rangle}}\right]\in\mathcal{S}(H_{\mathbf{R}}).
\end{equation}
In particular, the action of the encoding map on the maximally mixed reference state $\rho_0=\mathbb{I}_d/d$ is
\begin{equation}
    \mathcal{N}[\rho_0]=\frac{1}{kd}\sum_{ija}|i\rangle\langle j|_{\mathbf{R}}\frac{\langle ja |ia\rangle_{\mathbf{EOW}}}{N_a^2}.
\end{equation}
Here we defined $N_a:=\sqrt{\sum_i\frac{1}{k}\langle ia|ia\rangle_{\mathbf{EOW}}}$. 

The question of the information recovery is whether there is an approximate inverse of $\mathcal{N}$, recovering the original state $\rho$ from $\mathcal{N}[\rho]$. This is indeed the case when $k\gg de^{S(E)}$ corresponds to the late time after the Page time. In the following, we assume that late time, namely $1\leq d\ll k,~e^{S(E)}$.

\subsection{Relative Entropy Difference}
We first study the relative entropy difference for $\mathcal{N}$. Small relative entropy difference guarantees that the averaged rotated Petz map is an approximate recovery map for $\mathcal{N}$. We first review that the relative entropy difference is indeed small at 
$k\gg de^{S(E)}$. 

For simplicity, we consider encoding a pure state $|a\rangle\in H_{\text{code}}$ in the Hawking radiation. We fix the reference state as $\rho_0=\mathbb{I}_d/d$, and consider the relative entropy difference
\begin{equation}
    D(\sigma_a|\rho_0)-D(\mathcal{N}[\sigma_a]|\mathcal{N}[\rho_0]),
\end{equation}
where $\sigma_a:=|a\rangle\langle a|$. Since we have $D(\sigma_a|\rho_0)=\log d$, we only need to evaluate $D(\mathcal{N}[\sigma_a]|\mathcal{N}[\rho_0])$. This can be accomplished by employing the replica trick
\begin{equation}
    D(\mathcal{N}[\sigma_a]|\mathcal{N}[\rho_0])=\frac{\partial}{\partial n}\Big{|}_{n=1}
    \left[\text{Tr}[\mathcal{N}[\sigma_a]^{n}]-\text{Tr}[\mathcal{N}[\rho_0]^{n-1}\mathcal{N}[\sigma_a]]\right].
\end{equation}
We can explicitly write
\begin{eqnarray}
    \text{Tr}[\mathcal{N}[\rho_0]^{n-1}\mathcal{N}[\sigma_a]]
    &=&
    \sum_{1\leq i_0,\cdots,i_{n-1}\leq k}\sum_{1\leq a_1,\cdots,a_{n-1}\leq d}
    \frac{1}{(kd)^{n-1}}\nonumber\\
    &\times&
    \frac{\langle i_1a|i_0a\rangle_{\mathbf{EOW}}
    \langle i_0a_{n-1}|i_{n-1}a_{n-1}\rangle_{\mathbf{EOW}}   \cdots\langle i_2a_1|i_1a_1\rangle_{\mathbf{EOW}}}
    {N_{a_{n-1}}^2\cdots N_{a_1}^2N_a^2}.
\end{eqnarray}
\begin{figure}[t]
 \begin{center}
 \includegraphics[width=11cm,clip]{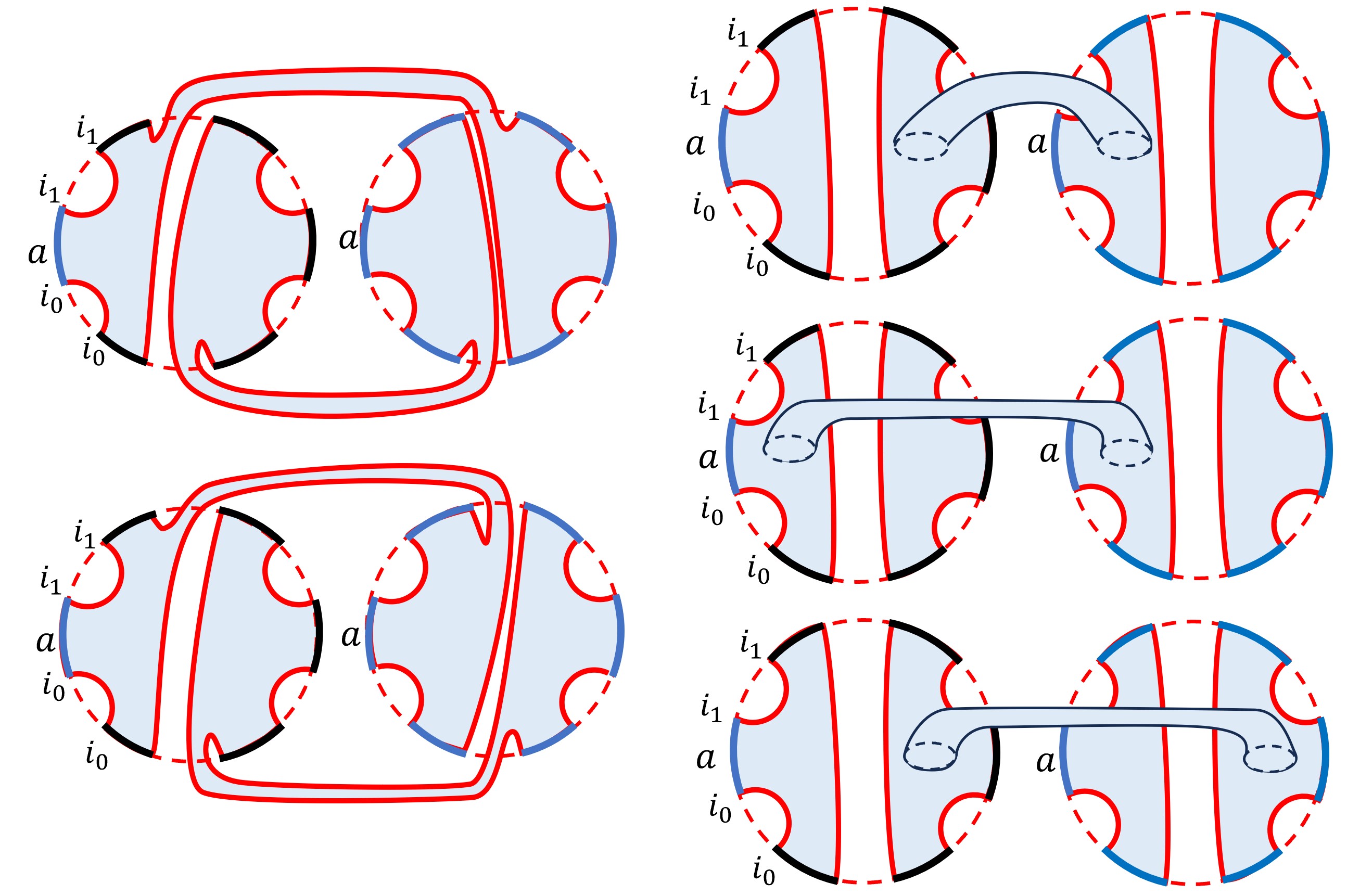}
 \end{center}
 \caption{Wormholes contributing to (\ref{eq:third}) at $k\gg de^{S(E)}$. The black and the blue lines correspond to the AdS boundaries. The black line corresponds to $\mathcal{N}[\rho_0]$, and the blue line corresponds to $\mathcal{N}[\sigma_a]$. The red lines are EOW branes. The dotted red lines are for index contractions of $i=1,\cdots,k$ and $a=1,\cdots, d$.}
 \label{fig:relative}
 \end{figure}
 The evaluation of this quantity can be done in the same way as the computation of the Page curve, replacing $e^{S_0}$ by $de^{S_0}$. Thus, by combining terms, the relative entropy difference is \cite{Vardhan:2021mdy}
\begin{equation}\label{eq:c1micro}
    D(\sigma_a|\rho_0)-D(\mathcal{N}[\sigma_a]|\mathcal{N}[\rho_0])
    =\left\{
    \begin{aligned}
    &\log d-\frac{k}{2e^{S(E)}}\left(1-\frac{1}{d}\right)+O(k^{-1})&  &(k< e^{S(E)})\\
    &\log \frac{de^{S(E)}}{k}-\frac{e^{S(E)}}{2k}+\frac{k}{2de^{S(E)}}+O(e^{-S(E)})&  &(e^{S(E)}<k<de^{S(E)}),\\
    &\frac{(d-1)e^{S(E)}}{2k}+O(e^{-S(E)})&  &(de^{S(E)}<k).
    \end{aligned}
    \right.
\end{equation}
In particular, at $k\gg de^{S(E)}$, the error in the encoding map $\mathcal{N}$ is small, and the Petz map is an approximate recovery map.

The main purpose of this paper is to understand whether this relative entropy difference is typically small in the ensemble. It suffices to study the fluctuation of the relative entropy $D(\mathcal{N}[\sigma_a]|\mathcal{N}[\rho_0])$. The fluctuation is given by
\begin{eqnarray}
    &&\left(\delta D(\mathcal{N}[\sigma_a]|\mathcal{N}[\rho_0])\right)^2
    \nonumber\\
    &=&
    \frac{\partial}{\partial n}\frac{\partial}{\partial m}\Big{|}_{n=m=1}\mathbb{E}\Big[
    \left(\text{Tr}[\mathcal{N}[\sigma_a]^{n}]-\text{Tr}[\mathcal{N}[\rho_0]^{n-1}\mathcal{N}[\sigma_a]]\right)
    \left(\text{Tr}[\mathcal{N}[\sigma_a]^{m}]-\text{Tr}[\mathcal{N}[\rho_0]^{m-1}\mathcal{N}[\sigma_a]]\right)\Big]\nonumber\\
    &-&
    \frac{\partial}{\partial n}\frac{\partial}{\partial m}\Big{|}_{n=m=1}\mathbb{E}\Big[
    \left(\text{Tr}[\mathcal{N}[\sigma_a]^{n}]-\text{Tr}[\mathcal{N}[\rho_0]^{n-1}\mathcal{N}[\sigma_a]]\right)\Big]
    \mathbb{E}\Big[ \left(\text{Tr}[\mathcal{N}[\sigma_a]^{m}]-\text{Tr}[\mathcal{N}[\rho_0]^{m-1}\mathcal{N}[\sigma_a]]\right)\Big].\nonumber\\
\end{eqnarray}
Here, we denote the ensemble average by $\mathbb{E}[\cdots]$. We compute this quantity by evaluating wormhole configurations at $k\gg de^{S(E)}$. The first contribution
\begin{equation}
    \partial_n\partial_m\Big{|}_{n=m=1}\Big(\mathbb{E}\Big[
    \text{Tr}[\mathcal{N}[\sigma_a]^{n}]
    \text{Tr}[\mathcal{N}[\sigma_a]^{m}]\Big]-\mathbb{E}\Big[
    \text{Tr}[\mathcal{N}[\sigma_a]^{n}]\Big]
    \mathbb{E}\Big[\text{Tr}[\mathcal{N}[\sigma_a]^{m}]\Big]\Big),
\end{equation}
is identical to the fluctuation of the entropy (\ref{eq:entropy}). The second contribution
\begin{equation}
    \partial_n\partial_m\Big{|}_{n=m=1}\Big(\mathbb{E}\Big[
    \text{Tr}[\mathcal{N}[\rho_0]^{n-1}\mathcal{N}[\sigma_a]]
    \text{Tr}[\mathcal{N}[\rho_0]^{m-1}\mathcal{N}[\sigma_a]]\Big]-\mathbb{E}\Big[
    \text{Tr}[\mathcal{N}[\rho_0]^{n-1}\mathcal{N}[\sigma_a]]\Big]
    \mathbb{E}\Big[\text{Tr}[\mathcal{N}[\rho_0]^{m-1}\mathcal{N}[\sigma_a]]\Big]\Big),
\end{equation}
is again identical to the fluctuation of the entropy (\ref{eq:entropy}) with $e^{S(E)}$ being replaced by $de^{S(E)}$. The third contribution
\begin{equation}\label{eq:third}
    -2\partial_n\partial_m\Big{|}_{n=m=1}
    \Big(\mathbb{E}\Big[
    \text{Tr}[\mathcal{N}[\rho_0]^{n-1}\mathcal{N}[\sigma_a]]
    \text{Tr}[\mathcal{N}[\sigma_a]^{m}]\Big]
    -\mathbb{E}\Big[
    \text{Tr}[\mathcal{N}[\rho_0]^{n-1}\mathcal{N}[\sigma_a]]\Big]
    \mathbb{E}\Big[\text{Tr}[\mathcal{N}[\sigma_a]^{m}]\Big]\Big),
\end{equation}
come from wormhole configuration described in Fig \ref{fig:relative} for $k\gg de^{S(E)}$. Summing over these contributions, we obtain
\begin{eqnarray}
    \delta D(\mathcal{N}[\sigma_a]|\mathcal{N}[\rho_0])
    &=&
    \frac{1}{2k}
    \sqrt{\frac{\log (e^{\frac{3}{2}}\frac{\Delta E}{a})}{\pi^2}(d-1)^2
    +(d-1)}+O\left(k^{-2}\right).
\end{eqnarray}
We see that when we compare with the relative entropy difference at $k\gg de^{S(E)}$, the fluctuation is suppressed by $e^{-S(E)}$. Thus, we conclude that the error in the encoding map has an exponentially suppressed fluctuation, and the approximate recovery map typically exists in the ensemble.


\subsection{Entanglement Fidelity of the Petz Recovery Map}

Next, we study the entanglement fidelity of the Petz map and examine possible errors in the recovery map. We first consider general quantum state $\rho=\sum_{ab}c_{ab}|a\rangle\langle b|$ on the code subspace. Before considering the entanglement fidelity, it is convenient to study the following amplitude
\begin{eqnarray}
    \langle a'|
    \Big(\mathcal{R}^{\text{Petz}}_{\mathcal{N},\rho_0}\circ\mathcal{N}\Big)
    \Big[\sum_{ab}c_{ab}|a\rangle\langle b|\Big]|b'\rangle    &=&
    \frac{1}{d}\text{Tr}\Big[(\mathcal{N}[|a'\rangle\langle b'|])^{\dagger}
    \mathcal{N}[\rho_0]^{-1/2}\mathcal{N}(\sum_{ab}c_{ab}|a\rangle\langle b|)\mathcal{N}[\rho_0]^{-1/2}\Big]\nonumber\\
    &=&
    \sum_{ab}c_{ab}\frac{1}{d}\text{Tr}\Big[(\mathcal{N}[|a'\rangle\langle b'|])^{\dagger}
    \mathcal{N}[\rho_0]^{-1/2}\mathcal{N}(|a\rangle\langle b|)\mathcal{N}[\rho_0]^{-1/2}\Big].\nonumber\\
\end{eqnarray}
It is clear that when this always gives $c_{a'b'}$, the Petz map is an exact recovery map. In the case of the PSSY model, the encoding map $\mathcal{N}$ has $U(d)$ symmetry on average; thus we can write
\begin{eqnarray}
    \mathbb{E}[\frac{1}{d}\text{Tr}\Big[(\mathcal{N}[|a'\rangle\langle b'|])^{\dagger}
    \mathcal{N}[\rho_0]^{-1/2}\mathcal{N}(|a\rangle\langle b|)\mathcal{N}[\rho_0]^{-1/2}\Big]]
    =
    c_1\delta_{aa'}\delta_{bb'}+c_2\frac{\delta_{ab}\delta_{a'b'}}{d}.
\end{eqnarray}
Then, the exact recovery is equivalent to $c_1=1,~c_2=0$. Since the composite map is trace-preserving, we generally have $c_1+c_2=1$. To evaluate $c_1$ and $c_2$, we can use the standard replica trick
\begin{eqnarray}
    F^{(n)}_{aba'b'}
    &:=&\frac{1}{d}\text{Tr}\Big[(\mathcal{N}[|a'\rangle\langle b'|])^{\dagger}
    \mathcal{N}[\rho_0]^{n}\mathcal{N}(|a\rangle\langle b|)\mathcal{N}[\rho_0]^{n}\Big]
    \nonumber\\
    &=&
    \frac{1}{d}\sum_{1\leq i_0,\cdots,j_0,\cdots\leq k}\sum_{1\leq a_1,\cdots,b_1,\cdots\leq d}
    \frac{1}{k^{2n+2}d^{2n}}
    \frac{\langle i_0b_n|j_nb_n\rangle_{\mathbf{EOW}}\cdots\langle j_2b_1|j_1b_1\rangle_{\mathbf{EOW}}
    \langle j_1b'|j_0a'\rangle_{\mathbf{EOW}}}
    {N_{b_n}^2\cdots N_{b_1}^2N_{b'}N_{a'}}
    \nonumber\\
    &\times&
    \frac{\langle j_0a_n|i_na_n\rangle_{\mathbf{EOW}}
    \cdots\langle i_2a_1|i_1a_1\rangle_{\mathbf{EOW}}
    \langle i_1a|i_0b\rangle_{\mathbf{EOW}}}{N_{a_n}^2\cdots N_{a_1}^2N_aN_b},
\end{eqnarray}
and take the limit $n\rightarrow -1/2$. Then, at large $k$ and $e^{S(E)}$, we have \cite{Penington:ReplicaWormholeWestCoast}
\begin{equation}\label{eq:c1micro}
    c_1=\left\{
    \begin{aligned}
    &\frac{k}{de^{S(E)}}+O(k^2e^{-2S(E)})&  (k&\ll de^{S(E)})\\
    & 1-\frac{de^{S(E)}}{4k}+O(e^{2S(E)}k^{-2})&  (k&\gg de^{S(E)}),
    \end{aligned}
    \right.
    c_2=\left\{
    \begin{aligned}
    &1-\frac{k}{de^{S(E)}}+O(k^2e^{-2S(E)})&  (k&\ll de^{S(E)})\\
    & \frac{de^{S(E)}}{4k}+O(e^{2S(E)}k^{-2})&  (k&\gg de^{S(E)}).
    \end{aligned}
    \right.
\end{equation}
Thus at $k\gg de^{S(E)}$, we have $c_1\approx 1$ annd $c_2\approx 0$. This means that the Petz map is an approximate recovery map on average. 

We now consider the entanglement fidelity. For a diagonal density matrix $\rho=\sum_{a}p_a|a\rangle\langle a|$, the entanglement fidelity of the channel $\mathcal{R}^{\text{Petz}}_{\mathcal{N},\rho_0}\circ\mathcal{N}$ is given by
\begin{eqnarray}
    F_e(\rho,~\mathcal{R}^{\text{Petz}}_{\mathcal{N},\rho_0}\circ\mathcal{N})&=&\langle\Phi_{\rho}|\Big{[}(\mathcal{R}^{\text{Petz}}_{\mathcal{N},\rho_0}\circ\mathcal{N})\otimes I_{d}\Big{]}\Big{[}|\Phi_\rho\rangle\langle\Phi_\rho|\Big{]}|\Phi_{\rho}\rangle
    \nonumber\\&=&
    \sum_{ab}p_ap_b F^{(-1/2)}_{abab}\nonumber\\
    &=&c_1+\frac{c_2}{d}\sum_ap_a^2,
\end{eqnarray}
here $|\Phi_{\rho}\rangle=\sum_ap^{1/2}_a|a\rangle|a\rangle$ is the purification of $\rho$. At $k\gg de^{S(E)}$, we have
\begin{eqnarray}\label{eq:EF}
    F_e(\rho,~\mathcal{R}^{\text{Petz}}_{\mathcal{N},\rho_0}\circ\mathcal{N})&=&1-\frac{de^{S(E)}}{4k}(1-\frac{1}{d}\sum_ap_a^2)+O(e^{2S(E)}k^{-2}).
\end{eqnarray}
In the following, we will consider the fluctuation of this entanglement fidelity.

\subsection*{Fluctuation in the Entanglement Fidelity of the Petz Map}

\begin{figure}[t]
 \begin{center}
 \includegraphics[width=11cm,clip]{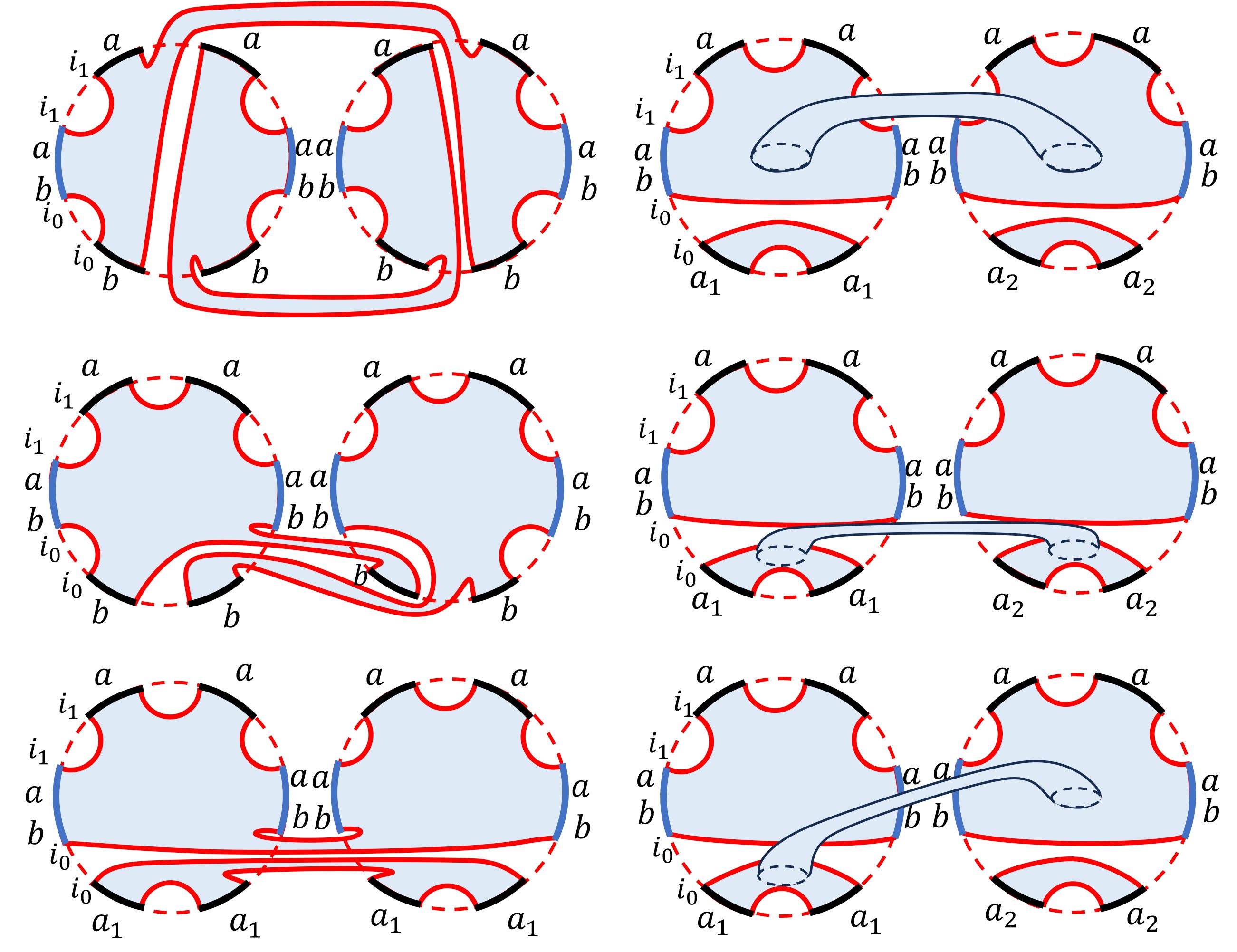}
 \end{center}
 \caption{Wormholes contributing to $f_1$ at $k\gg de^{S(E)}$. The black and the blue lines correspond to the AdS boundaries. The black line corresponds to $\mathcal{N}[\rho_0]$, and the blue line corresponds to $\mathcal{N}[|a\rangle\langle b|]$. The red lines are EOW branes. The dotted red lines are for index contractions of $i=1,\cdots,k$ and $a=1,\cdots, d$.}
 \label{fig:non}
 \end{figure}

\begin{figure}[t]
 \begin{center}
 \includegraphics[width=11cm,clip]{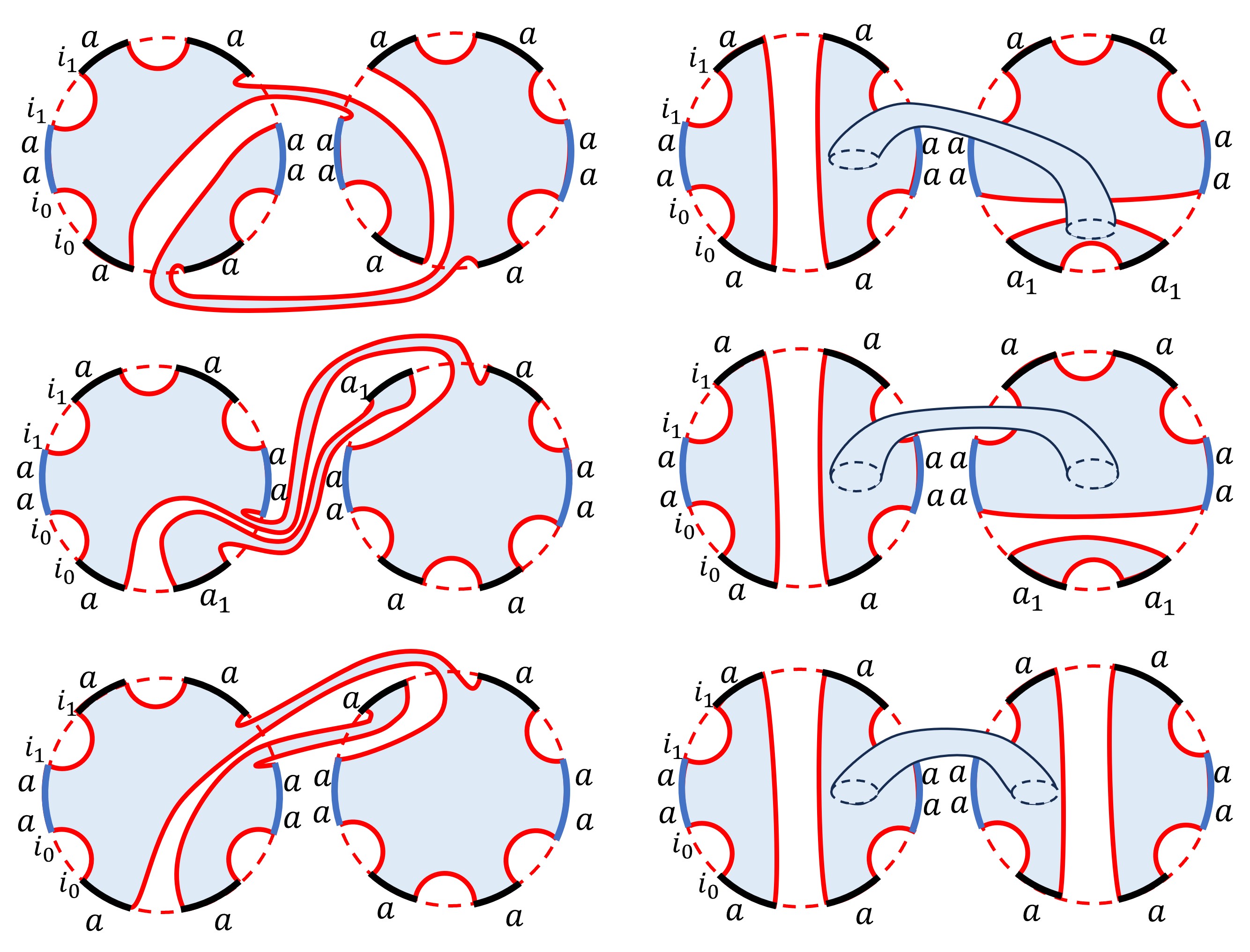}
 \end{center}
 \caption{Wormholes contributing to $f_2$ at $k\gg de^{S(E)}$. The black and the blue lines correspond to the AdS boundaries. The black line corresponds to $\mathcal{N}[\rho_0]$, and the blue line corresponds to $\mathcal{N}[|a\rangle\langle a|]$. }
 \label{fig:tube}
 \end{figure}

In order to evaluate the fluctuation of the entanglement fidelity, it is useful to evaluate the fluctuation of $F^{(-1/2)}_{abab}$. We will evaluate
\begin{eqnarray}
    \mathbb{E}[F^{(-1/2)}_{abab}F^{(-1/2)}_{abab}]-\mathbb{E}[F^{(-1/2)}_{abab}]\mathbb{E}[F^{(-1/2)}_{abab}]=f_1+\delta_{ab}f_2.
\end{eqnarray}
Showing both $f_1$ and $f_2$ are small at large $k$ implies that the Petz map is indeed an approximate recovery map typically.
The wormholes contributing to $f_1$ at $k\gg de^{S(E)}$ are described in Fig \ref{fig:non}.
The result is
\begin{equation}
    f_1=\frac{1}{32k^2}
    \left(4+d+2d^2
    \frac{\log (e^{\frac{3}{2}}\frac{\Delta E}{a})}{\pi^2}\right)+O(e^{S(E)}k^{-3}).
\end{equation}
The wormholes contributing to $f_2$ at $k\gg de^{S(E)}$ are described in Fig \ref{fig:tube}.
The result is
\begin{equation}
    f_2=\frac{1}{32k^2}\left(-6+d+(2-4d)\frac{\log (e^{\frac{3}{2}}\frac{\Delta E}{a})}{\pi^2}\right)+O(e^{S(E)}k^{-3}).
\end{equation}
Both $f_1$ and $f_1+f_2$ are non-negative for any $d$ as they should. 

Consider the case when the quantum state is pure $\sigma_a=|a\rangle\langle a|$. Then, the entanglement fidelity is
\begin{equation}\label{eq:Fe}
    F_e(\sigma_a,~\mathcal{R}^{\text{Petz}}_{\mathcal{N},\rho_0}\circ\mathcal{N})=F^{(-1/2)}_{aaaa}=1-\frac{de^{S(E)}}{4k}(1-\frac{1}{d})+O(e^{2S(E)}k^{-2}),
\end{equation}
and its fluctuation is 
\begin{equation}
    \delta F_e(\sigma_a,~\mathcal{R}^{\text{Petz}}_{\mathcal{N},\rho_0}\circ\mathcal{N})=\frac{1}{4k}\sqrt{(d-1)+(1-2d+d^2)\frac{\log(e^{\frac{3}{2}}\frac{\Delta E}{a})}{\pi^2}}+O(e^{S(E)}k^{-2}).
\end{equation}
Comparing with the average errors in (\ref{eq:Fe}), we see that the fluctuation is exponentially suppressed by $e^{-S(E)}$. Thus, we again conclude that the Petz map is typically an approximate recovery map at $k\gg de^{S(E)}$.

\subsection{Coherent Information Loss}\label{section:coherent}

In this subsection, we compute the coherent information loss and compare it with the entanglement fidelity of the Petz map (\ref{eq:EF}), using the inequality (\ref{eq:ineq}). Again, small coherent information loss implies the existence of an approximate recovery map.

Coherent information in the PSSY model was discussed in \cite{Balasubramanian:2022fiy}. We consider the case when the input density matrix is the maximally mixed state
$\rho=\sum_{a}d^{-1}|a\rangle\langle a|$. We deote its purification as $|\Phi_{\rho}\rangle=\sum_ad^{-1/2}|a\rangle|a\rangle$. Then the coherent information loss of $\mathcal{N}$ ad the state $\rho$ is given by
\begin{equation}
    \delta_c(\rho,\mathcal{N})=\log d
    -S\Big(\text{Tr}_{\bold{EOW}}\Big[\sum_{a}d^{-1}|\Psi_a\rangle\langle\Psi_a|\Big]\Big)
    +S\Big(\sum_{ab}d^{-1}|a\rangle\langle b|\text{Tr}_{\bold{EOW}}\Big[|\Psi_a\rangle\langle\Psi_b|\Big]\Big).
\end{equation}
We can evaluate this quantity easily by noticing that the second term is the entropy of the Hawking radiation with a replacement $e^{S(E)}\rightarrow de^{S(E)}$, and the third term is also the entropy of the Hawking radiation with a replacement $k\rightarrow dk$. Then, we can write the coherent information loss explicitly as
\begin{equation}\label{eq:coherent}
    \delta_c(\rho,\mathcal{N})
    =\left\{
    \begin{aligned}
    & 2\log d+\frac{k}{2de^{S(E)}}-\frac{dk}{2e^{S(E)}}+O(k^{-1})&  &(dk< e^{S(E)})\\
    & \log \frac{de^{S(E)}}{k}+\frac{k}{2de^{S(E)}}-\frac{e^{S(E)}}{2dk}+O(k^{-1})+O(e^{-S(E)})&  &(dk>e^{S(E)}~,k<de^{S(E)})\\
    & \frac{de^{S(E)}}{2k}(1-\frac{1}{d^2})+O(e^{-S(E)})&  &(de^{S(E)}<k).
    \end{aligned}
    \right.
\end{equation}
We see that at $k\gg de^{S(E)}$, the coherent information loss is small, guaranteeing the existence of an approximate recovery map.

Next, we evaluate the mutual information loss, which turns out to be very similar. For this purpose it suffices to evaluate $S\Big(\sum_{ab}d^{-1}|a\rangle\langle b|\langle\Psi_b|\Psi_a\rangle\Big)$. From $d\ll\text{Min}[k,~e^{S(E)}]$, it is given by
\begin{equation}
    S\Big(\sum_{ab}d^{-1}|a\rangle\langle b|\langle\Psi_b|\Psi_a\rangle\Big)=\log d+O(k^{-1})+O(e^{-S(E)}).
\end{equation}
Thus, the mutual information loss is 
\begin{equation}
    \delta_m(\rho,\mathcal{N})
    =\frac{1}{2}\delta_c(\rho,\mathcal{N})+O(k^{-1})+O(e^{-S(E)}).
\end{equation}
Substitution of this formula to (\ref{eq:mutual}) yields the identical formula as (\ref{eq:ineq}) for (\ref{eq:coherent}).

\section{Conclusion}

In this paper, we examined whether the information recovery from the black hole interior is typically possible in the ensemble of gravitational theories. Both the relative entropy difference and one minus the entanglement fidelity of the Petz map behave as $\propto de^{S(E)}/k$ at large $k$ on average, which guarantees the existence of an approximate recovery map of the interior information. We have shown that their fluctuations behave as $d/k$, which is exponentially suppressed by $e^{-S(E)}$ compared to $\propto de^{S(E)}/k$. This implies that information recovery is typically possible in the ensemble of theories.

We note that there are a number of important questions regarding the information recovery from the black hole interior, such as the origin of the ensemble averaging in quantum gravity. We have also not answered all the questions about the fluctuation of the recovery measures. We have essentially limited to the case of pure quantum state as the interior information. To evaluate the fluctuations for a more general input state, one needs to compute $\mathbb{E}[F^{(-1/2)}_{abab}F^{(-1/2)}_{cdcd}]$ in the case of the entanglement fidelity. We have only considered the maximally mixed state in the coherent information. It would be interesting to consider more general input states, as well as the fluctuations of the coherent information loss. We expect the fluctuations of these measures are similarly suppressed by $e^{-S(E)}$ compared to the signal, while detailed studies are required. 

It is also interesting that the characterization of the Petz map is under investigation. We note that the Petz map has better properties than the averaged rotated Petz map in terms of retrodiction \cite{Parzygnat:2022ldx}.


\paragraph{Acknowledgements}
We are grateful to F.~Buscemi for helpful comments and correspondence. We thank R.~Bousso for discussions. 





\bibliographystyle{JHEP}
\bibliography{Main.bib}

\providecommand{\href}[2]{#2}\begingroup\raggedright\begin{thebibliography}{10}

\bibitem{Bousso:2023efc}
R.~Bousso and M.~Miyaji, \emph{{Fluctuations in the Entropy of Hawking
  Radiation}},  \href{https://arxiv.org/abs/2307.13920}{{\ttfamily
  2307.13920}}.

\bibitem{Ryu:2006bv}
S.~Ryu and T.~Takayanagi, \emph{{Holographic derivation of entanglement entropy
  from AdS/CFT}},
  \href{https://doi.org/10.1103/PhysRevLett.96.181602}{\emph{Phys. Rev. Lett.}
  {\bfseries 96} (2006) 181602}
  [\href{https://arxiv.org/abs/hep-th/0603001}{{\ttfamily hep-th/0603001}}].

\bibitem{Hubeny:2007xt}
V.E.~Hubeny, M.~Rangamani and T.~Takayanagi, \emph{{A Covariant holographic
  entanglement entropy proposal}},
  \href{https://doi.org/10.1088/1126-6708/2007/07/062}{\emph{JHEP} {\bfseries
  07} (2007) 062} [\href{https://arxiv.org/abs/0705.0016}{{\ttfamily
  0705.0016}}].

\bibitem{Faulkner:2013ana}
T.~Faulkner, A.~Lewkowycz and J.~Maldacena, \emph{{Quantum corrections to
  holographic entanglement entropy}},
  \href{https://doi.org/10.1007/JHEP11(2013)074}{\emph{JHEP} {\bfseries 11}
  (2013) 074} [\href{https://arxiv.org/abs/1307.2892}{{\ttfamily 1307.2892}}].

\bibitem{Engelhardt:2014gca}
N.~Engelhardt and A.C.~Wall, \emph{{Quantum Extremal Surfaces: Holographic
  Entanglement Entropy beyond the Classical Regime}},
  \href{https://doi.org/10.1007/JHEP01(2015)073}{\emph{JHEP} {\bfseries 01}
  (2015) 073} [\href{https://arxiv.org/abs/1408.3203}{{\ttfamily 1408.3203}}].

\bibitem{Czech:2012bh}
B.~Czech, J.L.~Karczmarek, F.~Nogueira and M.~Van~Raamsdonk, \emph{{The Gravity
  Dual of a Density Matrix}},
  \href{https://doi.org/10.1088/0264-9381/29/15/155009}{\emph{Class. Quant.
  Grav.} {\bfseries 29} (2012) 155009}
  [\href{https://arxiv.org/abs/1204.1330}{{\ttfamily 1204.1330}}].

\bibitem{Almheiri:2014lwa}
A.~Almheiri, X.~Dong and D.~Harlow, \emph{{Bulk Locality and Quantum Error
  Correction in AdS/CFT}},
  \href{https://doi.org/10.1007/JHEP04(2015)163}{\emph{JHEP} {\bfseries 04}
  (2015) 163} [\href{https://arxiv.org/abs/1411.7041}{{\ttfamily 1411.7041}}].

\bibitem{Pastawski:2015qua}
F.~Pastawski, B.~Yoshida, D.~Harlow and J.~Preskill, \emph{{Holographic quantum
  error-correcting codes: Toy models for the bulk/boundary correspondence}},
  \href{https://doi.org/10.1007/JHEP06(2015)149}{\emph{JHEP} {\bfseries 06}
  (2015) 149} [\href{https://arxiv.org/abs/1503.06237}{{\ttfamily
  1503.06237}}].

\bibitem{Jafferis:2015del}
D.L.~Jafferis, A.~Lewkowycz, J.~Maldacena and S.J.~Suh, \emph{{Relative entropy
  equals bulk relative entropy}},
  \href{https://doi.org/10.1007/JHEP06(2016)004}{\emph{JHEP} {\bfseries 06}
  (2016) 004} [\href{https://arxiv.org/abs/1512.06431}{{\ttfamily
  1512.06431}}].

\bibitem{Dong:2016eik}
X.~Dong, D.~Harlow and A.C.~Wall, \emph{{Reconstruction of Bulk Operators
  within the Entanglement Wedge in Gauge-Gravity Duality}},
  \href{https://doi.org/10.1103/PhysRevLett.117.021601}{\emph{Phys. Rev. Lett.}
  {\bfseries 117} (2016) 021601}
  [\href{https://arxiv.org/abs/1601.05416}{{\ttfamily 1601.05416}}].

\bibitem{Hayden:2016cfa}
P.~Hayden, S.~Nezami, X.-L.~Qi, N.~Thomas, M.~Walter and Z.~Yang,
  \emph{{Holographic duality from random tensor networks}},
  \href{https://doi.org/10.1007/JHEP11(2016)009}{\emph{JHEP} {\bfseries 11}
  (2016) 009} [\href{https://arxiv.org/abs/1601.01694}{{\ttfamily
  1601.01694}}].

\bibitem{Harlow:2016vwg}
D.~Harlow, \emph{{The Ryu\textendash{}Takayanagi Formula from Quantum Error
  Correction}}, \href{https://doi.org/10.1007/s00220-017-2904-z}{\emph{Commun.
  Math. Phys.} {\bfseries 354} (2017) 865}
  [\href{https://arxiv.org/abs/1607.03901}{{\ttfamily 1607.03901}}].

\bibitem{Cotler:2017erl}
J.~Cotler, P.~Hayden, G.~Penington, G.~Salton, B.~Swingle and M.~Walter,
  \emph{{Entanglement Wedge Reconstruction via Universal Recovery Channels}},
  \href{https://doi.org/10.1103/PhysRevX.9.031011}{\emph{Phys. Rev. X}
  {\bfseries 9} (2019) 031011}
  [\href{https://arxiv.org/abs/1704.05839}{{\ttfamily 1704.05839}}].

\bibitem{Chen:2019gbt}
C.-F.~Chen, G.~Penington and G.~Salton, \emph{{Entanglement Wedge
  Reconstruction using the Petz Map}},
  \href{https://doi.org/10.1007/JHEP01(2020)168}{\emph{JHEP} {\bfseries 01}
  (2020) 168} [\href{https://arxiv.org/abs/1902.02844}{{\ttfamily
  1902.02844}}].

\bibitem{Barnum}
H.~Barnum and E.~Knill, \emph{{Reversing quantum dynamics with near-optimal
  quantum and classical fidelity}}, {\emph{Journal of Mathematical Physics}
  {\bfseries 43} (2002) 2097}.

\bibitem{Penington:2019npb}
G.~Penington, \emph{{Entanglement Wedge Reconstruction and the Information
  Paradox}}, \href{https://doi.org/10.1007/JHEP09(2020)002}{\emph{JHEP}
  {\bfseries 09} (2020) 002}
  [\href{https://arxiv.org/abs/1905.08255}{{\ttfamily 1905.08255}}].

\bibitem{Almheiri:2019psf}
A.~Almheiri, N.~Engelhardt, D.~Marolf and H.~Maxfield, \emph{{The entropy of
  bulk quantum fields and the entanglement wedge of an evaporating black
  hole}}, \href{https://doi.org/10.1007/JHEP12(2019)063}{\emph{JHEP} {\bfseries
  12} (2019) 063} [\href{https://arxiv.org/abs/1905.08762}{{\ttfamily
  1905.08762}}].

\bibitem{Almheiri:IslandFormula}
A.~Almheiri, R.~Mahajan, J.~Maldacena and Y.~Zhao, \emph{{The Page curve of
  Hawking radiation from semiclassical geometry}},
  \href{https://doi.org/10.1007/JHEP03(2020)149}{\emph{JHEP} {\bfseries 03}
  (2020) 149} [\href{https://arxiv.org/abs/1908.10996}{{\ttfamily
  1908.10996}}].

\bibitem{Penington:ReplicaWormholeWestCoast}
G.~Penington, S.H.~Shenker, D.~Stanford and Z.~Yang, \emph{{Replica wormholes
  and the black hole interior}},
  \href{https://doi.org/10.1007/JHEP03(2022)205}{\emph{JHEP} {\bfseries 03}
  (2022) 205} [\href{https://arxiv.org/abs/1911.11977}{{\ttfamily
  1911.11977}}].

\bibitem{Almheiri:ReplicaWormholeEastCoast}
A.~Almheiri, T.~Hartman, J.~Maldacena, E.~Shaghoulian and A.~Tajdini,
  \emph{{Replica Wormholes and the Entropy of Hawking Radiation}},
  \href{https://doi.org/10.1007/JHEP05(2020)013}{\emph{JHEP} {\bfseries 05}
  (2020) 013} [\href{https://arxiv.org/abs/1911.12333}{{\ttfamily
  1911.12333}}].

\bibitem{Vardhan:2021mdy}
S.~Vardhan, J.~Kudler-Flam, H.~Shapourian and H.~Liu, \emph{{Mixed-state
  entanglement and information recovery in thermalized states and evaporating
  black holes}}, \href{https://doi.org/10.1007/JHEP01(2023)064}{\emph{JHEP}
  {\bfseries 01} (2023) 064}
  [\href{https://arxiv.org/abs/2112.00020}{{\ttfamily 2112.00020}}].

\bibitem{Balasubramanian:2022fiy}
V.~Balasubramanian, A.~Kar, C.~Li and O.~Parrikar, \emph{{Quantum error
  correction in the black hole interior}},
  \href{https://doi.org/10.1007/JHEP07(2023)189}{\emph{JHEP} {\bfseries 07}
  (2023) 189} [\href{https://arxiv.org/abs/2203.01961}{{\ttfamily
  2203.01961}}].

\bibitem{Akers:2022qdl}
C.~Akers, N.~Engelhardt, D.~Harlow, G.~Penington and S.~Vardhan, \emph{{The
  black hole interior from non-isometric codes and complexity}},
  \href{https://arxiv.org/abs/2207.06536}{{\ttfamily 2207.06536}}.

\bibitem{Czech:2023rbh}
B.~Czech, S.~Shuai and H.~Tang, \emph{{Information recovery in the
  Hayden-Preskill protocol}},
  \href{https://arxiv.org/abs/2310.16988}{{\ttfamily 2310.16988}}.

\bibitem{Nakayama:2023kgr}
Y.~Nakayama, A.~Miyata and T.~Ugajin, \emph{{The Petz (lite) recovery map for
  scrambling channel}},  \href{https://arxiv.org/abs/2310.18991}{{\ttfamily
  2310.18991}}.

\bibitem{Hayden:2007cs}
P.~Hayden and J.~Preskill, \emph{{Black holes as mirrors: Quantum information
  in random subsystems}},
  \href{https://doi.org/10.1088/1126-6708/2007/09/120}{\emph{JHEP} {\bfseries
  09} (2007) 120} [\href{https://arxiv.org/abs/0708.4025}{{\ttfamily
  0708.4025}}].

\bibitem{Hayden:2018khn}
P.~Hayden and G.~Penington, \emph{{Learning the Alpha-bits of Black Holes}},
  \href{https://doi.org/10.1007/JHEP12(2019)007}{\emph{JHEP} {\bfseries 12}
  (2019) 007} [\href{https://arxiv.org/abs/1807.06041}{{\ttfamily
  1807.06041}}].

\bibitem{Maldacena:2004rf}
J.M.~Maldacena and L.~Maoz, \emph{{Wormholes in AdS}},
  \href{https://doi.org/10.1088/1126-6708/2004/02/053}{\emph{JHEP} {\bfseries
  02} (2004) 053} [\href{https://arxiv.org/abs/hep-th/0401024}{{\ttfamily
  hep-th/0401024}}].

\bibitem{Saad:2018bqo}
P.~Saad, S.H.~Shenker and D.~Stanford, \emph{{A semiclassical ramp in SYK and
  in gravity}},  \href{https://arxiv.org/abs/1806.06840}{{\ttfamily
  1806.06840}}.

\bibitem{Saad:2019pqd}
P.~Saad, \emph{{Late Time Correlation Functions, Baby Universes, and ETH in JT
  Gravity}},  \href{https://arxiv.org/abs/1910.10311}{{\ttfamily 1910.10311}}.

\bibitem{Harlow:2018tqv}
D.~Harlow and D.~Jafferis, \emph{{The Factorization Problem in
  Jackiw-Teitelboim Gravity}},
  \href{https://doi.org/10.1007/JHEP02(2020)177}{\emph{JHEP} {\bfseries 02}
  (2020) 177} [\href{https://arxiv.org/abs/1804.01081}{{\ttfamily
  1804.01081}}].

\bibitem{Bousso:2019ykv}
R.~Bousso and M.~Toma\v{s}evi\'c, \emph{{Unitarity From a Smooth Horizon?}},
  \href{https://doi.org/10.1103/PhysRevD.102.106019}{\emph{Phys. Rev. D}
  {\bfseries 102} (2020) 106019}
  [\href{https://arxiv.org/abs/1911.06305}{{\ttfamily 1911.06305}}].

\bibitem{Bousso:2020kmy}
R.~Bousso and E.~Wildenhain, \emph{{Gravity/ensemble duality}},
  \href{https://doi.org/10.1103/PhysRevD.102.066005}{\emph{Phys. Rev. D}
  {\bfseries 102} (2020) 066005}
  [\href{https://arxiv.org/abs/2006.16289}{{\ttfamily 2006.16289}}].

\bibitem{Kretschmann_2004}
D.~Kretschmann and R.F.~Werner, \emph{Tema con variazioni: quantum channel
  capacity}, \href{https://doi.org/10.1088/1367-2630/6/1/026}{\emph{New Journal
  of Physics} {\bfseries 6} (2004) 26}.

\bibitem{5429118}
F.~Buscemi and N.~Datta, \emph{The quantum capacity of channels with
  arbitrarily correlated noise},
  \href{https://doi.org/10.1109/TIT.2009.2039166}{\emph{IEEE Transactions on
  Information Theory} {\bfseries 56} (2010) 1447}.

\bibitem{Junge:2015lmb}
M.~Junge, R.~Renner, D.~Sutter, M.M.~Wilde and A.~Winter, \emph{{Universal
  Recovery Maps and Approximate Sufficiency of Quantum Relative Entropy}},
  \href{https://doi.org/10.1007/s00023-018-0716-0}{\emph{Annales Henri
  Poincare} {\bfseries 19} (2018) 2955}
  [\href{https://arxiv.org/abs/1509.07127}{{\ttfamily 1509.07127}}].

\bibitem{schumacher2002}
B.~Schumacher and M.D.~Westmoreland, \emph{Approximate quantum error
  correction}, \href{https://doi.org/10.1023/A:1019653202562}{\emph{Quantum
  Information Processing} {\bfseries 1} (2002) 5}.

\bibitem{PhysRevA.77.012309}
F.~Buscemi, \emph{Entanglement measures and approximate quantum error
  correction}, \href{https://doi.org/10.1103/PhysRevA.77.012309}{\emph{Phys.
  Rev. A} {\bfseries 77} (2008) 012309}.

\bibitem{FB}
F.~Buscemi, \emph{Irreversibility of entanglement loss},  in \emph{Theory of
  Quantum Computation, Communication, and Cryptography}, Y.~Kawano and
  M.~Mosca, eds., (Berlin, Heidelberg), pp.~16--28, Springer Berlin Heidelberg,
  2008.

\bibitem{Teitelboim:1983ux}
C.~Teitelboim, \emph{{Gravitation and Hamiltonian Structure in Two Space-Time
  Dimensions}}, \href{https://doi.org/10.1016/0370-2693(83)90012-6}{\emph{Phys.
  Lett. B} {\bfseries 126} (1983) 41}.

\bibitem{Jackiw:1984je}
R.~Jackiw, \emph{{Lower Dimensional Gravity}},
  \href{https://doi.org/10.1016/0550-3213(85)90448-1}{\emph{Nucl. Phys. B}
  {\bfseries 252} (1985) 343}.

\bibitem{Maldacena:2016upp}
J.~Maldacena, D.~Stanford and Z.~Yang, \emph{{Conformal symmetry and its
  breaking in two dimensional Nearly Anti-de-Sitter space}},
  \href{https://doi.org/10.1093/ptep/ptw124}{\emph{PTEP} {\bfseries 2016}
  (2016) 12C104} [\href{https://arxiv.org/abs/1606.01857}{{\ttfamily
  1606.01857}}].

\bibitem{Stanford:2017thb}
D.~Stanford and E.~Witten, \emph{{Fermionic Localization of the Schwarzian
  Theory}}, \href{https://doi.org/10.1007/JHEP10(2017)008}{\emph{JHEP}
  {\bfseries 10} (2017) 008}
  [\href{https://arxiv.org/abs/1703.04612}{{\ttfamily 1703.04612}}].

\bibitem{Yang:2018gdb}
Z.~Yang, \emph{{The Quantum Gravity Dynamics of Near Extremal Black Holes}},
  \href{https://doi.org/10.1007/JHEP05(2019)205}{\emph{JHEP} {\bfseries 05}
  (2019) 205} [\href{https://arxiv.org/abs/1809.08647}{{\ttfamily
  1809.08647}}].

\bibitem{Saad:2019lba}
P.~Saad, S.H.~Shenker and D.~Stanford, \emph{{JT gravity as a matrix
  integral}},  \href{https://arxiv.org/abs/1903.11115}{{\ttfamily 1903.11115}}.

\bibitem{Stanford:2019vob}
D.~Stanford and E.~Witten, \emph{{JT gravity and the ensembles of random matrix
  theory}}, \href{https://doi.org/10.4310/ATMP.2020.v24.n6.a4}{\emph{Adv.
  Theor. Math. Phys.} {\bfseries 24} (2020) 1475}
  [\href{https://arxiv.org/abs/1907.03363}{{\ttfamily 1907.03363}}].

\bibitem{Parzygnat:2022ldx}
A.J.~Parzygnat and F.~Buscemi, \emph{{Axioms for retrodiction: achieving
  time-reversal symmetry with a prior}},
  \href{https://doi.org/10.22331/q-2023-05-23-1013}{\emph{Quantum} {\bfseries
  7} (2023) 1013} [\href{https://arxiv.org/abs/2210.13531}{{\ttfamily
  2210.13531}}].

\end{thebibliography}\endgroup
\end{document}